\documentclass[superscriptaddress,twocolumn, showpacs,aps,prl]{revtex4-1}
\usepackage{natbib}
\usepackage[utf8]{inputenc}
\usepackage{amsmath,amssymb,color}
\usepackage{graphicx}
\usepackage{latexsym}
\usepackage{multirow}
\newcommand{\basilisk}{{\usefont{T1}{pzc}{m}{n}Basilisk}}

\begin{document}

\title{The natural breakup length of a steady capillary jet}
\author{ Alfonso M. Ga\~n\'an-Calvo}
\affiliation{Departamento de Ingenier{\'\i}a Aerospacial y Mec\'anica de Fluidos,
Universidad de Sevilla.\\
Camino de los Descubrimientos s/n 41092, Spain.}
\author{Henry N. Chapman}
\affiliation{Center for Free-Electron Laser Science, Deutsches Elektronen Synchrotron (DESY), Notkestrasse 85, 22607 Hamburg, Germany.}
\affiliation{Dept. Physics, University of Hamburg, Luruper Chaussee 149, 22761 Hamburg, Germany.}
\affiliation{Centre for Ultrafast Imaging, Luruper Chaussee 149, 22761 Hamburg, Germany.}
\author{Michael Heymann}
\affiliation{Max Planck Institute of Biochemistry, Am Klopferspitz 18, 82152 Martinsried, Germany}
\author{Max O. Wiedorn}
\affiliation{Center for Free-Electron Laser Science, Deutsches Elektronen Synchrotron (DESY), Notkestrasse 85, 22607 Hamburg, Germany.}
\affiliation{Dept. Physics, University of Hamburg, Luruper Chaussee 149, 22761 Hamburg, Germany.}
\affiliation{Centre for Ultrafast Imaging, Luruper Chaussee 149, 22761 Hamburg, Germany.}
\author{Juraj Knoska}
\affiliation{Center for Free-Electron Laser Science, Deutsches Elektronen Synchrotron (DESY), Notkestrasse 85, 22607 Hamburg, Germany.}
\affiliation{Dept. Physics, University of Hamburg, Luruper Chaussee 149, 22761 Hamburg, Germany.}
\author{Yang Du}
\affiliation{Center for Free-Electron Laser Science, Deutsches Elektronen Synchrotron (DESY), Notkestrasse 85, 22607 Hamburg, Germany.}
\author{Braulio Ga\~nan-Riesco}
\affiliation{Ingeniatrics Tec. S.L., 41900 Camas, Sevilla, Spain.}
\author{Miguel A. Herrada}
\affiliation{Departamento de Ingenier{\'\i}a Aerospacial y Mec\'anica de Fluidos,
Universidad de Sevilla.\\
Camino de los Descubrimientos s/n 41092, Spain.}
\author{Jos{\'e} M. L{\'o}pez-Herrera}
\affiliation{Departamento de Ingenier{\'\i}a Aerospacial y Mec\'anica de Fluidos,
Universidad de Sevilla.\\
Camino de los Descubrimientos s/n 41092, Spain.}
\author{Francisco Cruz-Mazo}
\affiliation{Departamento de Ingenier{\'\i}a Aerospacial y Mec\'anica de Fluidos,
Universidad de Sevilla.\\
Camino de los Descubrimientos s/n 41092, Spain.}
\affiliation{Department of Mechanical and Aerospace Engineering, Princeton University, Princeton, NJ 08544, USA..\\
Camino de los Descubrimientos s/n 41092, Spain.}
\author{Sa\v{s}a Bajt}
\affiliation{Deutsches Elektronen Synchrotron (DESY), Notkestra\ss{}e 85, 22607 Hamburg, Germany.}
\author{Jos\'e M. Montanero}
\affiliation{Departmento de Ingenier\'{\i}a Mec\'anica, Energ\'etica y de los Materiales and\\
Instituto de Computaci\'on Cient\'{\i}fica Avanzada (ICCAEx),\\
Universidad de Extremadura, Avda.\ de Elvas s/n, E-06071 Badajoz, Spain}

\begin{abstract}
Despite their fundamental and applied importance, a general model to predict the natural breakup length of steady capillary jets has not been proposed yet. In this work, we derive a scaling law with two universal constants to calculate that length as a function of the liquid properties and operating conditions. These constants are determined by fitting the scaling law to a large set of experimental and numerical measurements, including previously published data. Both the experimental and numerical jet lengths conform remarkably well to the proposed scaling law. This law is explained in terms of the growth of perturbations excited by the jet breakup itself.
\end{abstract}

\pacs{47.55.D-, 47.55.db, 47.55.df}

\maketitle

The shape, instability, and breakup of capillary jets have attracted scientific curiosity since long ago \cite{P49,R79a}. Capillary jets provide a gentle and reproducible way to transport a liquid without solid contact at distances from a source long compared to the source transversal size \cite{EV08}. The demand for precise means to deliver tiny liquid samples has explosively grown with the advent of faster and more sensitive detection and analysis procedures. For example, maximum jet length combined with jet diameters below 5 $\mu$m is desirable in serial femtosecond crystallography (SFX) \citep{Cetal11,W18}, which is one of the major applications for microjets.

Despite the numerous studies of jet instabilities \citep{KLAS03,GG08,GG09,GGCC14}, we are still lacking a general theoretical model for predicting the breakup (intact) length of steadily and freely released capillary jets \citep{G98a,GPG01,DWSWSSD08,SLYY09,SLYY10,GHM14,ZBS18}. The breakup length can be estimated from the classical temporal linear stability analysis of the jet. In this case, one assumes that the most unstable (dominant) temporal mode is responsible for the breakup. This mode is supposed to be triggered by a perturbation next to the jet inception region and is convected by the jet, which implies that the residence time in the jet scales as the inverse of the dominant mode growth rate. Rayleigh's theory for inviscid cylindrical capillary jets \citep{R79a} and its subsequent refinements to account for different factors allows one to calculate the growth rate of the dominant mode, which leads to a scaling law for the breakup length. However, the temporal linear stability analysis presents two important drawbacks: (i) the initial perturbation amplitude is a free parameter, which implies that the prefactor of the jet length scaling cannot be predicted; and (ii) the model does not contemplate the energy feedback coming from the jet breakup, as will be described below. Using the temporal stability analysis, \citet{IYXS18} derived two scalings for the breakup length of jets under the action of an axial electric field in the limits of small and large Reynolds numbers. Despite the drawbacks mentioned above, good agreement between those scalings and experimental data was found.

An accurate study of the natural (not externally excited) capillary breakup of jets has been conducted by \citet{U16}, who described the breakup mechanism as a ``self-destabilizing loop''. In this loop, the energy of the perturbations responsible for the breakup comes exclusively from earlier breakup events. The idea is that the excess of interfacial energy after the interface pinch-off feeds the perturbations that cause the subsequent breakup. In fact, the surface of the drop is, on average, about 20\% smaller than that of the jet portion which produces that drop. Although most of the excess of interfacial energy is eventually dissipated in the droplets, a small component of the energy spectral density propagates upstream and form imperceptible perturbations. These perturbations grow until causing the next breakup. In this work, we make use of this view of the problem to rationalize our experimental observations.

\begin{figure}[htb]
\centering
\includegraphics[width=0.47\textwidth]{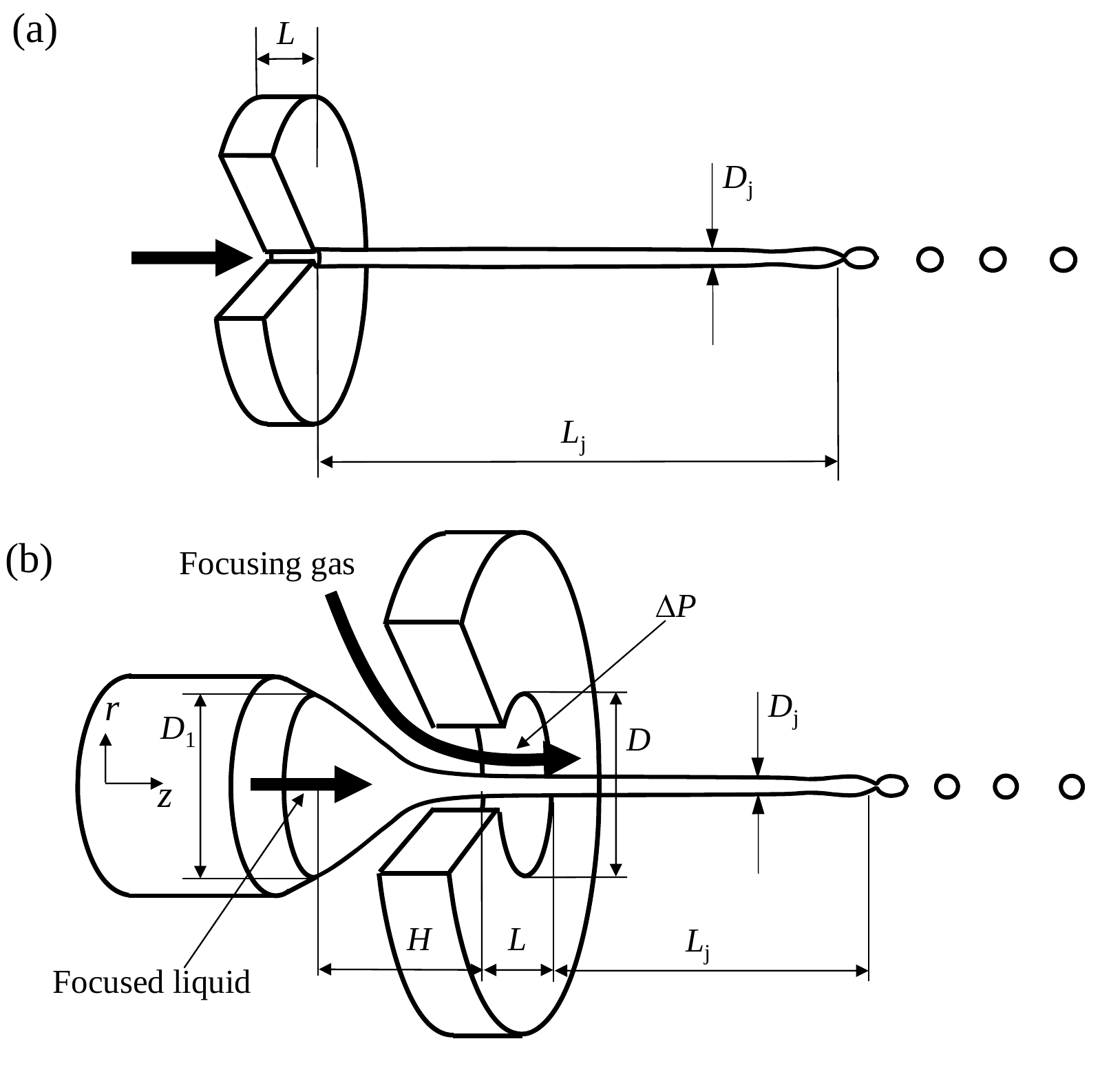}
\vspace{-3mm}
\caption{Ballistic (a) and flow-focused (b) capillary jet emitted from an orifice of diameter $D$. In flow focusing, $\Delta P$ is the pressure drop along the gas streamlines in the discharge orifice.}
\label{fig1}\vspace{-2mm}
\end{figure}

We aim at showing the universality and determinism of the natural breakup mechanism for capillary jets in the absence of body forces. For this purpose, we consider both ballistic and flow-focused \citep{G98a} microjets (Fig.\ \ref{fig1}). In a ballistic jet, the pressure applied to the liquid reservoir is essentially transformed into kinetic energy to overcome the resistance offered by surface tension to the jet formation at the orifice exit. In gaseous flow focusing \citep{G98a}, the jet is accelerated driven by both the pressure and viscous forces exerted by a high-speed outer stream in the discharge orifice. This way of focusing the energy necessary for the jet emission allows the reduction of the jet diameter $D_j$ below that of the discharge orifice $D$.

The ballistic and flow focusing configurations can be characterized in terms of the Weber and Capillary numbers
\begin{equation}
\text{We}=\frac{{\cal D}_{G}}{d_\sigma}\quad \text{and} \quad \text{Ca}=\left(\frac{\mu^2 \Delta P}{\sigma^2\rho}\right)^{1/2},
\end{equation}
where $d_\sigma\equiv\sigma/\Delta P$ and ${\cal D}_{G}\equiv\left[8\rho Q^2/(\pi^2 \Delta P)\right]^{1/4}$. The Weber and Capillary numbers are defined in terms of the liquid density $\rho$, viscosity $\mu$ and surface tension $\sigma$, as well as the pressure $\Delta P$ applied to produce the jet and the emitted flow rate $Q$. They do not involve any geometrical parameter. In the ballistic configuration, the diameter ${\cal D}_{G}$ is practically the same as that of jet, $D_j$. In flow focusing, $D_j<{\cal D}_{G}$ due to the viscous diffusion of momentum from the faster gaseous stream to the jet, which increases the jet speed $V_j$.

We focus on the Rayleigh regime in which the jet breaks up axisymmetrically due to surface tension. This regime has technological relevance because it leads to longer jets and more monodisperse collections of droplets. It is typically obtained for We$\gtrsim 1$ and $\text{We}_g=\rho_g (V_j-V_g)^2 D_j/(2\sigma)\lesssim 0.2$, where $\rho_g$ and $V_g$ are the gaseous environment density and velocity, respectively \citep{EV08}. In fact, Weber numbers We smaller than unity generally lead to dripping, while asymmetric perturbations produce shorter jets and more irregular breakup for We$_g\gtrsim 0.2$ \citep{RG08}. As will be seen, although ballistic and flow-focused jets are emitted in a very different way, the breakup length $L_j$ follows the same scaling law.

\begin{figure}[htb]
\centering
\includegraphics[width=0.5\textwidth]{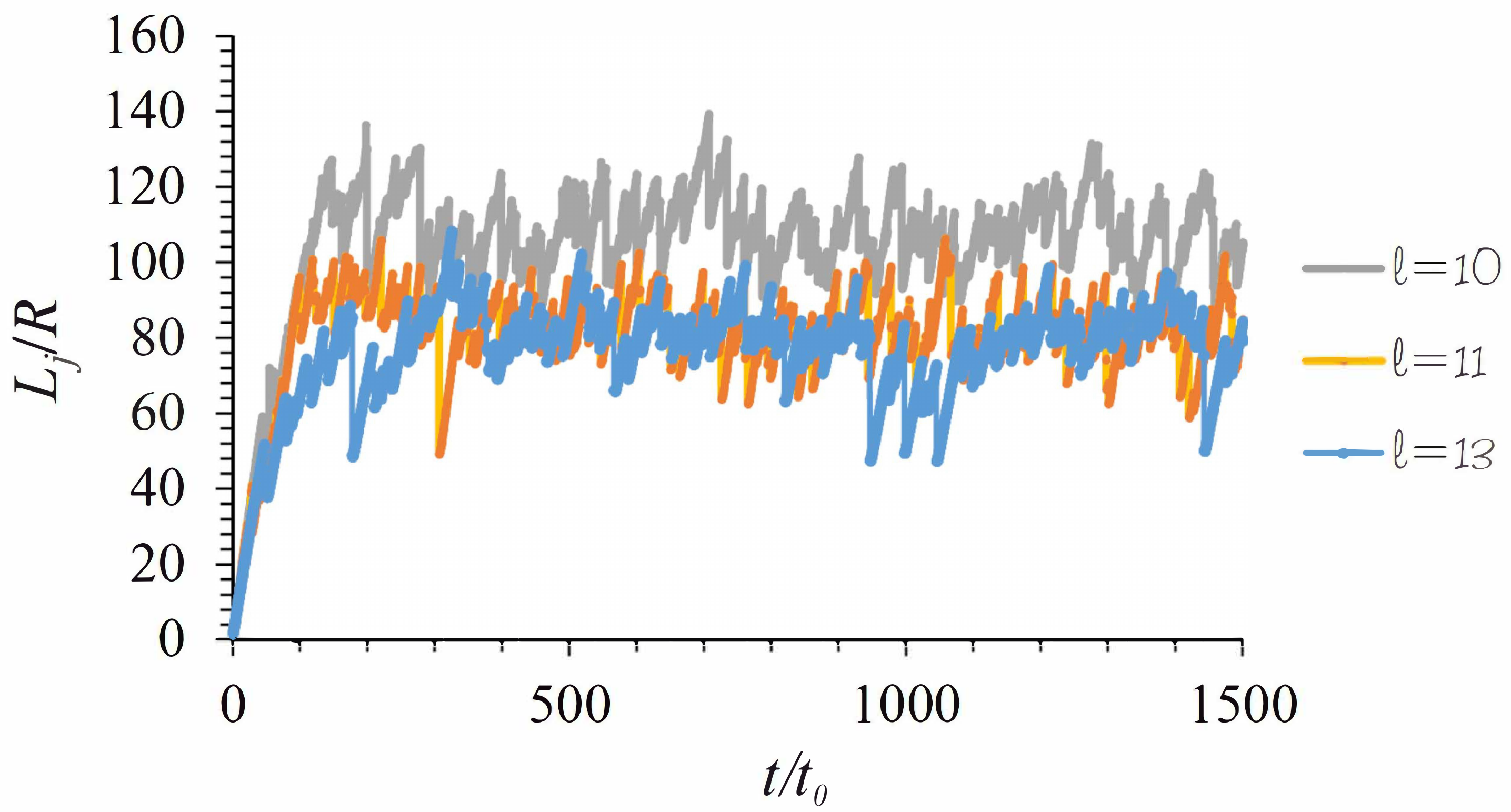}\\
\vspace{-3mm}
\caption{Distance $L_j(t)$ of the jet front position from the feeding tube exit as a function of time for different levels $\ell=$10, 11, and 13 of spatial discretization (see the Supplemental Material). The time is measured in terms of the capillary time $t_0=(\rho R^3/\sigma)^{1/2}$, where $R$ is the tube radius. The density and viscosity of the gas environment are 1000 and 100 times smaller than those of the liquid domain, respectively.}
\label{figSM2}
\end{figure}

In spite of significant efforts made in this field \citep{KLAS03,GG08,GG09,GGCC14}, it has not as yet been established why capillary jets that break up spontaneously become destabilized in a rather deterministic manner at a well-located position, within relatively narrow statistical limits. For the sake of illustration, Fig.\ \ref{figSM2} shows the evolution of the front of a ballistic jet injected at $t=0$ with We=5 and Ca=$5\times10^{-3}$ and surrounded by a dynamically negligible environment. The simulation was performed using the free software \basilisk \, \citep{Basilisk}. More details of the numerical simulation can be found in the Supplemental Material. As can be observed, the jet spontaneously breaks up at a relatively constant distance $L_j$ from the orifice (the standard deviation is around 13\% of the average value).

When droplets are formed from the jet breakage, about 80\% of the surface energy flowing towards the breakup region leaves the jet in the form of spherical droplets. The excess of surface energy splits into two parts after the breakup: one stays in the drop, provoking oscillations that are eventually dissipated by viscosity, and the other gets trapped on the jet side. The latter part propagates in the upstream direction, which is the only route available because the (continuous) jet domain ends at the breakup region. According to \citet{U16}, a small amount of this energy is dissipated by viscosity in the jet, while the rest feeds the growth of the perturbation responsible for the next breakup event.

\begin{figure}[htb]
\centering
\includegraphics[width=0.5\textwidth]{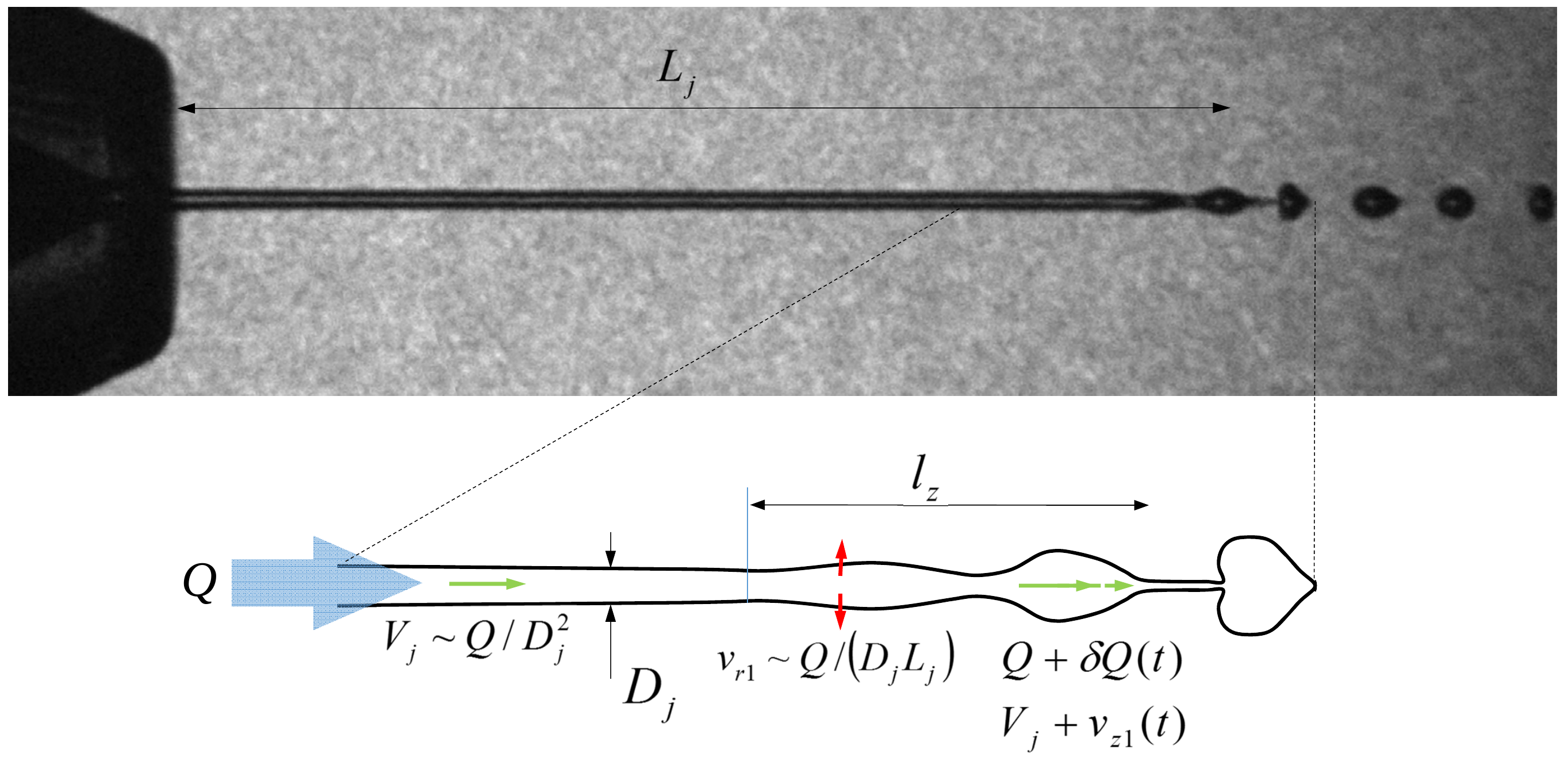}\\
\vspace{-3mm}
\caption{Scaling of the perturbation in the breakup region.}
\label{figS2}
\end{figure}

The liquid incompressibility implies that the axial distance $l_z$ from the breakup point along which the jet velocity $V_j$ is perturbed (Fig.\ \ref{figS2}) verifies $v_{z1}/l_z\sim v_{r1}/D_j$, where $v_{r1}$ and $v_{z1}$ represent the radial and axial perturbation velocity, repectively. Due to the convective character \citep{HM90a} of the jetting regime, the perturbation produced by the breakup travels only a few jet diameters in the upstream direction, i.e., $l_z\sim D_j$, which implies that $v_{r1}\sim v_{z1}$. This means that the radial and axial kinetic energies are commensurate with each other, $\rho v_{r1}^2\sim \rho v_{z1}^2$, and the radial and axial viscous stresses are also of the same order of magnitude, $\mu v_{r1}/D_j\sim \mu v_{z1}/l_z$. These results allow us to retain only the radial kinetic energy and viscous stress in the balance of energy described above.

The production rate of the energy released following the droplet formation scales as $Q\sigma/D_j$. The production rate of the kinetic energy associated with the radial perturbation scales as $Q\rho v_{r1}^2$, while the rate of energy dissipated by viscosity in the jet scales as $Q\mu v_{r1}/D_j$. Taking into account these scalings, the balance of energy described above leads to the expression
\begin{equation}
\label{e0}
\frac{\sigma}{D_j}=a_{\rho}\, \rho v_{r1}^2+a_{\mu}\, \mu\frac{v_{r1}}{D_j},
\end{equation}
where $a_{\rho}$ and $a_{\mu}$ are dimensionless constants. This analysis considers the energy balance in the breakup region. Therefore, one may expect the constants $a_{\rho}$ and $a_{\mu}$ not to depend on the procedure used to emit the jet.

An important element of the present analysis is the scale $v_{r1}$ of the perturbation velocity. As is similarly done in the temporal stability analysis \citep{IYXS18}, we assume that the residence time in the jet scales as the time for the transversal perturbation to cause the free surface pinching, i.e., $L_j/V_j\sim D_j/v_{r1}$. Since $D_j\sim l_z$ and $v_{r1}\sim v_{z1}$, we get $l_z/L_j\sim v_{z1}/V_j$, which implies that the perturbation caused by the breakup propagates upstream a distance much smaller than the jet length. This constitutes an essential difference with respect to the scenario assumed in temporal linear stability analysis, in which perturbations grow from the jet inception. The fact that the jet breakup can be regarded as a local phenomenon in terms of the jet length can be considered as the defining condition of the jetting regime. In fact, if $l_z$ were commensurate with $L_j$, then the growth of the pertubation would be affected by the discharge orifice, as it is characteristic of the dripping regime.

Taking into account the above scaling for $v_{r1}$, Eq.\ (\ref{e0}) becomes
\begin{equation}
\frac{\sigma}{D_j} =b_{\rho}\, \rho \left( \frac{Q}{D_j L_j}\right)^2 +b_{\mu}\, \mu\frac{Q}{D_j L_j}D_j^{-1}.
\label{secondorder}
\end{equation}
This result can be expressed in dimensionless form as
\begin{equation}
\frac{L_j}{d_\sigma}=\alpha_{\rho} \zeta, \quad \zeta=\text{We}^2\left[\left(\text{We}+\alpha_{\mu}^2\text{Ca}^2\right)^{1/2}-\alpha_{\mu}\text{Ca}\right]^{-1},
\label{ef1}
\end{equation}
where $\alpha_{\rho}=\left(b_{\rho}\pi^2/8\right)^{1/2}$ and $\alpha_{\mu}=b_{\mu}/(2b_{\rho}^{1/2})$ are dimensionless constants. The constant $\alpha_{\rho}^{-1}$ reflects the fraction of the surface energy that feeds the transversal motion in the breakup region, while $\alpha_{\mu}$ is an indicator of the ratio of the viscous to the kinetic energies. For low-viscosity jets ($\alpha_{\mu}\text{Ca}\ll \text{We}^{1/2}$), Eq.\ (\ref{ef1}) reduces to
\begin{equation}
\frac{L_j}{d_\sigma}\simeq \alpha_{\rho}\text{We}^{3/2}\left(1+\alpha_{\mu}\text{Oh}\right),
\end{equation}
where $\text{Oh}=\text{Ca}/\text{We}^{1/2}$ is the Ohnesorge number.

To validate Eq.\ (\ref{ef1}), we make use of (i) 30 measurements of the length of ballistic jets emanating from an orifice 250 $\mu$m in diameter, (ii) 400 measurements of the length of flow-focused jets using four liquids (see Table \ref{tab1}) and six ejectors (see Table \ref{tab2}), (iii) 10 and 20 measurements taken from Ref.\ \citep{U16} of the length of ballistic jets emanating from an orifice 0.4 and 1 mm in diameter, respectively, and (iv) the diameters calculated in two direct numerical simulations of ballistic jets.

\begin{table*}
\begin{tabular}{|c|c|c|c|}
\hline liquid & $\rho$ (kg$\cdot$m$^{-3}$) & $\sigma$ (N$\cdot$m$^{-1}$) & $\mu$ (Pa$\cdot$s) \\ \hline
water (22$^o$C) & 1000 & 0.072 & 0.001 \\
water/ethanol (65/35 v/v \%) (20$^o$C) & 943 & 0.035 & 0.0026 \\
ethanol (22$^o$C) & 795 & 0.023 & 0.00125 \\
water/glycerol (20/80 v/v \%) (22$^o$C) & 1217 & 0.065 & 0.0914 \\
\hline
\end{tabular}
\caption{Properties of the liquids used in experiments. }
\label{tab1}
\end{table*}

\begin{table}
\begin{tabular}{|c|c|c|c|}
\hline ejector& orifice shape & dimensions ($\mu$m) & $D_1$ ($\mu$m) \\
\hline
1 & slit & 15$\times$45 & 30 \\
2 & slit & 20$\times$60 & 30 \\
3 & round & 30 & 30 \\
4 & round & 50 & 50 \\
5 & round & 75 & 75 \\
6 & round & 70 & 100 \\
\hline
\end{tabular}
\caption{Micronozzles used in the flow focusing experiments. The third column indicates the dimensions of the discharge orifice. Ejector 6 was described by \citet{BAHKKWCB15} and \citet{ZBS18}.}
\label{tab2}
\end{table}

\begin{figure}[htb]
\centering
\includegraphics[width=0.35\textwidth]{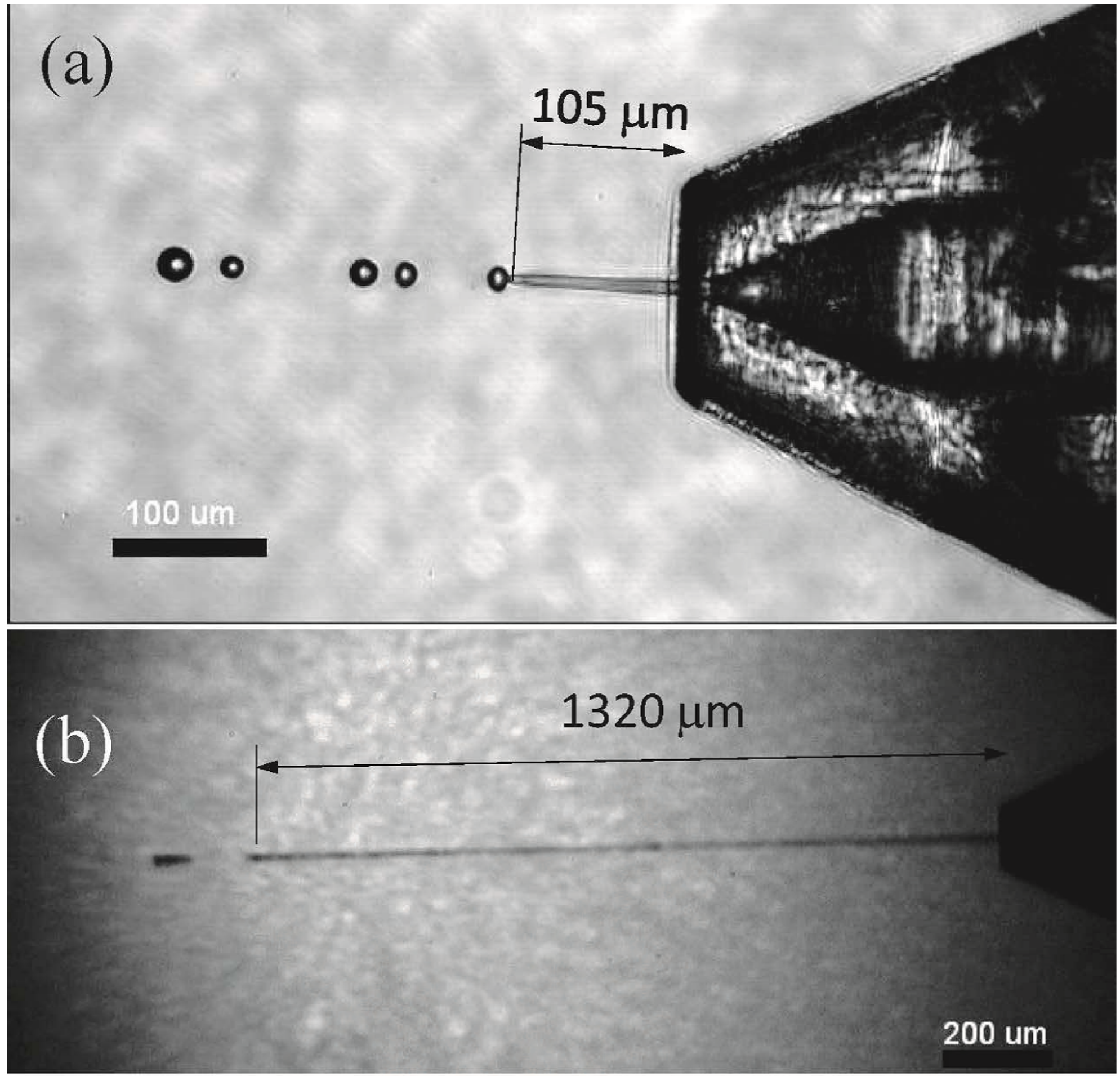}
\vspace{-2mm}
\caption{(a) A short water microjet emitted with $Q=8.2$ $\mu$l/min and $G_o=10.4$ mg/min. The focused capillary meniscus from whose apex the jet issues can be observed through the translucent plastic nozzle. The jet length is approximately 105 $\mu$m. (b) A long microjet of a mixture of water/glycerol (20/80 v/v \%) emitted with $Q=20$ $\mu$l/min and $G_o=7$ mg/min. In the two experiments, the jet was emitted with Ejector 4.}
\label{fig5}\vspace{-0.2cm}
\end{figure}

In our experiments with ballistic jets, the liquid was injected with a precision syringe pump into the air at the atmospheric pressure through an orifice of diameter $D=250$ $\mu$m. In the flow focusing experiments, the liquid jet was focused with helium. The liquid flow rate $Q$ and gas mass flow rate $G_o$ were fixed using a precision syringe pump and a GP1 (Equilibar Inc., Fletcher, USA) gas pressure regulator, respectively. The gas mass flow rate was monitored using a Bronkhorst flow meter. The pressure drop through the orifice, $\Delta P$, was calculated by assuming adiabatic flow with a discharge coefficient of $\eta=0.85$ (see the Supplemental Material). This value was obtained as the average over thousands of measurements with orifice sizes from 20 $\mu$m to 2 mm. We jetted into both atmospheric pressure and rough vacuum. In all the experiments conducted in this work, images of the jet were acquired with a high-speed video camera. The jet was illuminated with a laser with a pulse duration between 5 and 500 ns (Fig.\ \ref{fig5}). The jet length was determined by visually pinpointing the average location of the breakup events during some seconds. We estimate that the magnitude of the breakup length fluctuations is 10\%-20\% of the average value, as also occurs in the simulations (Fig.\ \ref{figSM2}).

\begin{figure}[htb]
\centering
\includegraphics[width=0.4\textwidth]{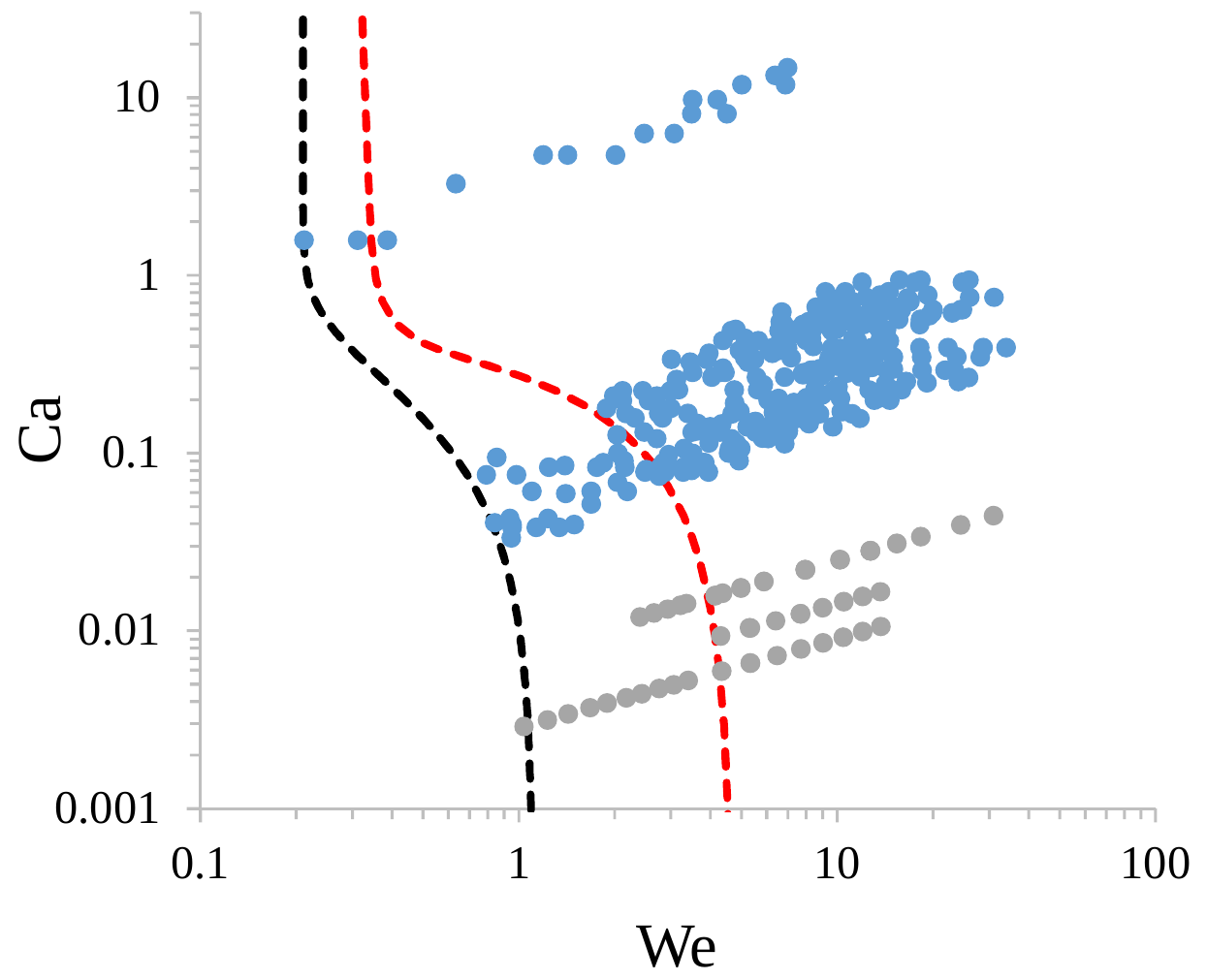}
\vspace{-2mm}
\caption{Values of We and Ca in our experiments. The grey and blue symbols correspond to ballistic and flow-focused jets, respectively. The red line corresponds to the convective-to-absolute instability transition for a cylindrical capillary jet in a vacuum \citep{LG86a}. The black line is a guide to the eye. The upper cloud of points corresponds to the liquid with the highest viscosity.}
\label{fig6}\vspace{-0.2cm}
\end{figure}

We measured jet diameters and lengths for conditions between the onset of dripping and asymmetric breakup. Figure \ref{fig6} shows the values of We and Ca in our experiments. Two and four orders of magnitude in We and Ca have been explored, respectively. Interestingly, the minimum length occurs for Weber numbers below the classical prediction of \citet{LG86a}. In flow focusing, this is the result of the well-known stabilizing effect of the coflowing gas stream. In both ballistic and flow-focused jets, the boundary layer growing on the inner side of the free surface of low-viscosity jets also delays the convective-to-absolute instability transition \citep{GHM14}. The maximum Weber number for axisymmetric breakup was around 35 for both ballistic and flow-focused jets. This upper limit corresponds to the maximum jet length obtained in the experiments, and it is reached for $Q\simeq 1.36\times 10^3\, Q_\sigma$, where $Q_\sigma\equiv [\sigma^4/(\rho \Delta P^3)]^{1/2}$ \cite{GM09}.

\begin{figure}[htb]
\centering
\includegraphics[width=0.5\textwidth]{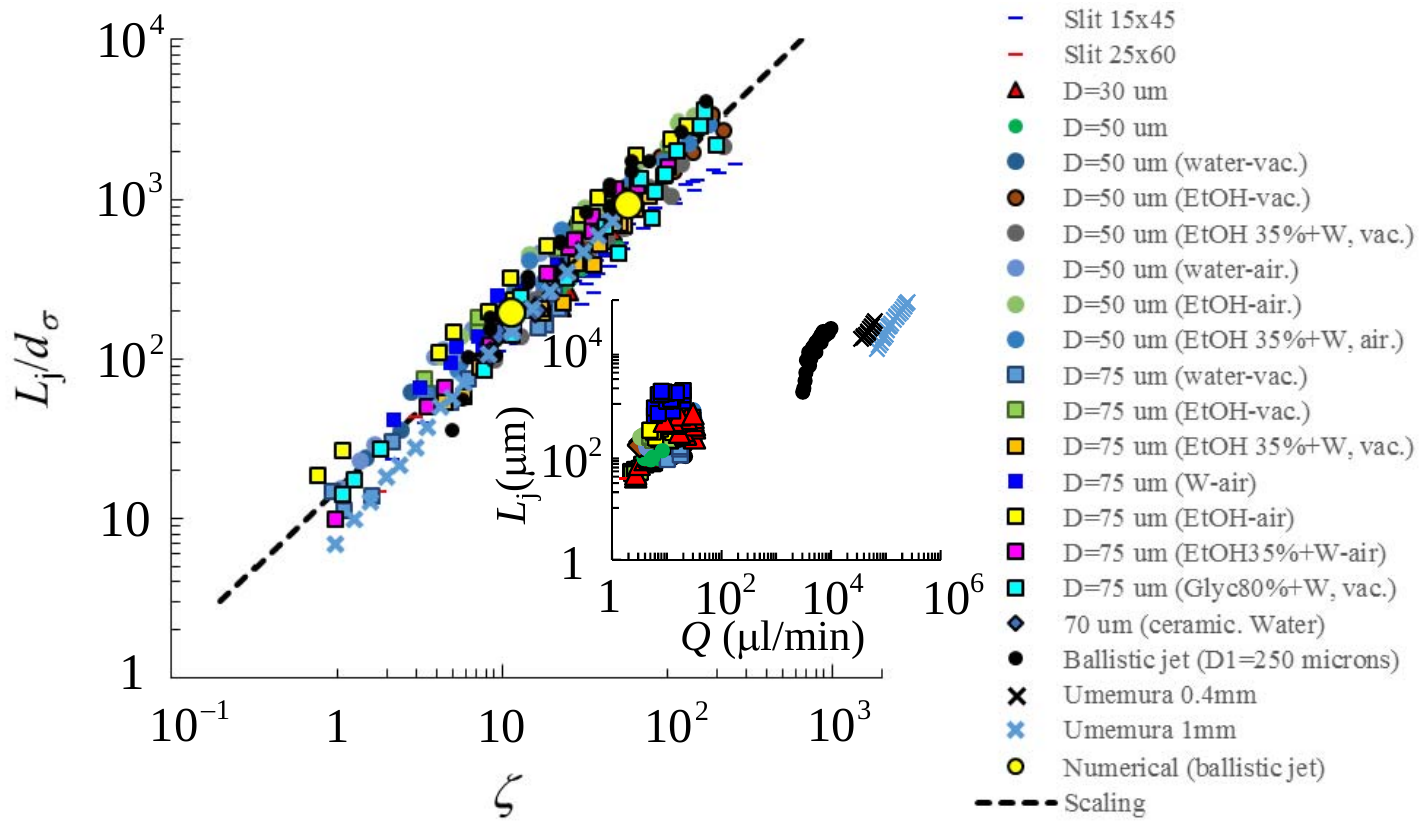}\\
\vspace{-4mm}
\caption{Breakup length $L_j/d_\sigma$ as a function of $\zeta$ obtained from both experiments and numerical simulations (symbols) and calculated from (\ref{ef1}) (line). The legend indicates the discharge orifice dimensions, the liquid, and the environment (vacuum or air). The inset shows $L_j$ as a function of $Q$.}
\label{fig3}\vspace{-0.2cm}
\end{figure}

\begin{figure}[htb]
\centering
\includegraphics[width=0.47\textwidth]{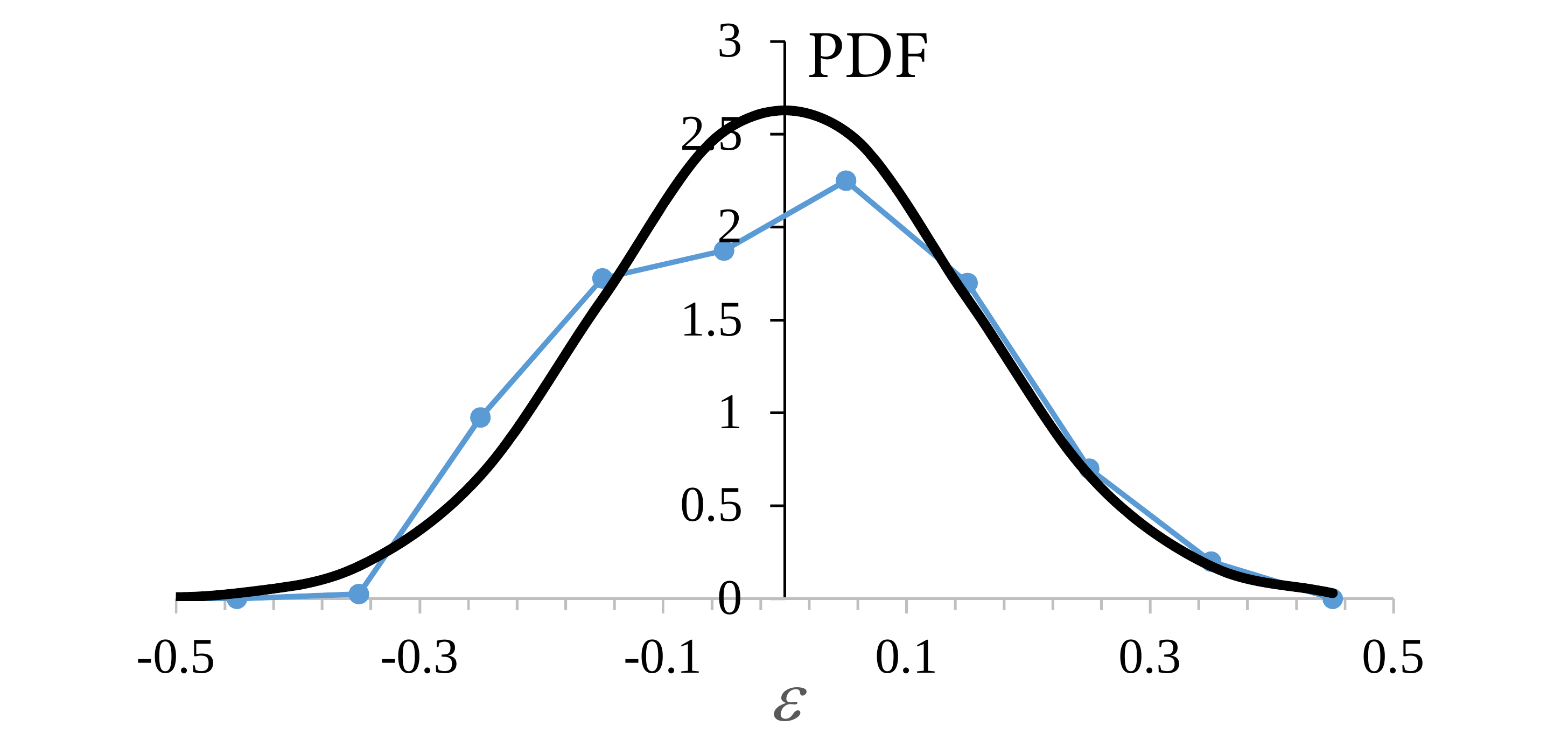}
\vspace{-0.2cm}
\caption{Probability density function of the logarithmic errors around the scaling law (\ref{ef1}) for $\alpha_{\rho}=15.015$ and $\alpha_{\mu}=0.53$ (symbols), and the corresponding normal distribution with zero average and variance equal to 0.0225 (line).}
\label{fig4}\vspace{-0.2cm}
\end{figure}

The best fit (\ref{ef1}) to all the data considered in this work is represented in Fig.\ \ref{fig3}. To obtain this fit, we calculated the probability density function PDF of the logarithmic error $\varepsilon=\log[\alpha_{\rho}\zeta(\alpha_{\rho},\alpha_{\mu})]-\log(L_j/d_\sigma)$ for different values of $\alpha_{\rho}$ and $\alpha_{\mu}$. A normal distribution with zero average is fitted to that function (Fig.\ \ref{fig4}). The optimum values of the constants $\alpha_{\rho}$ and $\alpha_{\mu}$ are those leading to the normal distribution with minimum variance. The minimum variance $s^2=0.0225$ was obtained for $\alpha_{\rho}=15.015$ and $\alpha_{\mu}=0.53$. The agreement of the resulting scaling law (\ref{ef1}) with both experiments and numerical simulations is remarkable. The relatively large value of $\alpha_{\rho}$ indicates the relatively small fraction of surface energy necessary to feed the upstream perturbations.

Equation (\ref{ef1}) exhibits remarkable agreement with experimental and numerical data for optimum values of $\alpha_{\rho}$ and $\alpha_{\mu}$ within a large parameter space and using different geometries. Using this model, one can select the best combination of liquid properties, nozzle outlet area and operating parameters to obtain the desired jet length, diameter and velocity for specific applications. This is remarkably valuable for designing and operating microjet devices in SFX \cite{W18,W18b}, where liquid formulations should also be compatible with protein crystals buffer solution (see the Supplemental Material).

As occurs in the classical temporal stability model, the breakup mechanism assumed in this work relies on the growth of the most unstable capillary mode, which explains the size of the produced droplets and the distance between them. The question is where and how this mode is excited. While the temporal stability approach assumes that the dominant capillary mode is triggered in the jet birth, the model adopted here supposes that this mode is excited in the breakup region. In most of the well-controlled experimental realizations, the jet breaks up in a rather deterministic manner in a well-located position. It has not as yet been elucidated why the jet birth could constitute a regular source of perturbation that explains that experimental fact. On the contrary, the jet breakup is a quasi-periodic source of energy which may regularly feed the perturbations leading to each breakup event. On the other hand, the temporal stability approach may suggest that the breakup length should depend on the details of the ejection procedure and geometry, which are expected to play a relevant role in the excitation of the dominant capillary mode. However, the experimental and numerical results presented here indicate that the breakup length essentially depends on the liquid properties ($\rho$, $\mu$ and $\sigma$) and operating parameters ($\Delta P$ and $Q$), which reinforces the idea that the perturbation origin is not located in the ejector. In any case, the model adopted here does not exclude the possibility that the temporal stability approach may satisfactorily explain the breakup length measurements under certain experimental conditions.

This work rationalizes experimental observations about the breakup length of steady capillary jets spontaneously breaking in the absence of significant body forces and interaction with the fluid environment. A quantitative theoretical analysis is still required to fully understand the mechanism that determines the breakup length.

We would like to thank Luigi Adriano (DESY) for technical support. This work was supported by the Ministerio de Econom{\'\i}a y Competitividad (Spain) through the project DPI2016-78887, and the German Science Foundation (DFG) through the Gottfried Wilhelm Leibniz program. MH acknowledges support from the Joachim Herz Foundation through an Add-on Fellowship. F. C-M. acknowledges support through the project funded by the European Union's 2020 Research and Innovation Program under the Marie Sklodowska-Curie Grant Agreement No 838997.

\end{document}